\documentclass{article}
\usepackage{amsmath}
\usepackage[english]{babel}
\usepackage[latin1]{inputenc}

\title{Cosmology: a bird's eye view}
\author{Alan A. Coley, Sigbj{\o}rn Hervik, Woei Chet Lim\\
{\it  \small Dalhousie University, Dept. of Mathematics and Statistics,}\\
{\it  \small Halifax, NS, Canada B3H 3J5}\\
\small E-mails: {\tt\small aac, herviks, wclim@mathstat.dal.ca}}
\begin{document}
\maketitle
\begin{abstract}
In this essay we discuss the difference in views of the Universe as seen by two different observers. While one of the observers follows a geodesic congruence defined by the geometry of the cosmological model, the other observer follows the fluid flow lines of a perfect fluid with a linear equation of state. We point out that the information these observers collect regarding the state of the Universe can be radically different; while one observes a non-inflating ever-expanding ever-lasting universe, the other observer can experience a dynamical behaviour reminiscent to that of quintessence or even that of a phantom cosmology leading to a 'big rip' singularity within finite time (but without the need for exotic forms of matter). 
\end{abstract}

\pagebreak
When Einstein was formulating the special theory and the
general theory of relativity, he realised the importance of
observers. In the formulation of the principle of Relativity and the
principle of Equivalence, upon which the general theory of relativity
is founded, the role of observers is of uttermost importance. Whenever
experiments or observerations are made our measurements necessarily
depend on our world-line. In this Essay we will continue in the spirit
of Einstein and point out the role the fundamental observer has for
our perception of the Universe.

In cosmology it is essential to specify a set of observers, or rather,
a congruence of world-lines from which experiments and observations
are made. However, quantities like the Hubble parameter depend on the choice of these
congruences and consequently these quantities will depend on the
observer. From the special theory of relativity, we are familiar with
the 'twin-paradox' for which two twins age differently because the
world-lines they follow are not identical.  As in the 'twin-paradox',
we will consider a Universe with two different observers (or rather two different
congruences of observers) which we will call Hugin and
Munin\footnote{In Norse mythology, Hugin and Munin are a pair of
  ravens which travel the world bearing news and information to the
  god Odin.}. The perception of the Universe as observed by Hugin and
Munin will depend on the paths they take. We will see that their
interpretation of the Universe can be radically different.

In our examples, we will consider a spatially homogeneous universe
containing a perfect fluid and we will assume that this fluid is not
necessarily co-moving with the spatially homogeneous
hypersurfaces \cite{KingEllis}. These models are prime examples of the type phenomenon we are interested in. For
these models there are
two naturally defined time-like congruences; namely, the vector field
orthogonal to the surfaces of transitivity, $n^{\mu}$, and the fluid
flow-lines, $u^{\mu}$. Whenever $n^{\mu}$ is not parallell to $u^{\mu}$ (which we shall assume in this Essay)  we will say that the fluid is \emph{tilted}. Moreover, we will let Hugin follow the congruence
$n^{\mu}$ while Munin follows the fluid flow-lines, $u^{\mu}$. In this
way, Hugin takes a 'geometric' point of view, while Munin follows the
fluid and views the Universe as seen from the fluid. These 
models have the advantage that they are dynamical solutions to
Einstein's field equations and can therefore model some features of
our real Universe.

The properties of the congruences that Hugin and Munin are following
can be very different. Hugin's path is necessarily geodesic,
vorticity-free and accelera\-tion-free. Munin's path, on the other hand,
does not need to be geodesic and can have both vorticity and
acceleration. As a consequence  the view  of the
Universe and its interpretation can also be very different. For example,
using solutions to the field equations, we will point out that while Hugin observes an
ever-expanding non-inflating universe, Munin can observe an inflating
universe with Hubble expansion similar to models with a quintessence fluid. In
other cases, while Hugin experiences an everywhere regular
ever-expanding universe, Munin will experience a universe that ends in
a violent 'big rip' singularity.

The question of which of these observers is the most 'physical one' is
a subtle one. If the observers are falling freely and not under the
influence of an external force then they will move along geodesics:
this point of view is represented by Hugin. However, this is an ideal
situation where the observer is moving in a 'void' in which there is
nothing that can exert pressure on him. If we consider the observer in
the real Universe and necessarily a part of the energy-momentum
tensor (i.e., the right-hand side of the Einstein equations) and
thereby a part of the source of the gravitational field, it is more
natural to take a matter point of view as represented by
Munin. This question is also related to which side of the Einstein equations are the most fundamental; the left-hand side, representing the geometry and the curvature, or the right-hand side, representing the matter and energy. 

Leaving this question aside, let us consider the consequence of the choice of an observer for some
real and physically interesting  solutions to the  Einstein field
equations. We consider the spatially homogeneous Bianchi type VIII model, which is a particularly suitable
model for illustrations. This model is the most general of the Bianchi
models and has sufficient richness to illustrate our points. We
will assume that the universe model possesses a tilted perfect fluid
with equation of state $p=(\gamma-1)\rho$ where $\gamma$ is a
constant. Of special interest are the values $\gamma=1$ (dust) and
$\gamma=4/3$ (radiation). First we consider the more common choice
where we see the Universe from Hugin's perspective. (All behaviours
referred to are the asymptotic behaviours at late times.)

For any $2/3<\gamma<2$, Hugin's record of events is as follows. Using dynamical systems methods \cite{DS1} it can be shown that \cite{HLim}:
\begin{itemize}
\item{} The Universe is ever-expanding, non-inflating and
  well-behaved into the future. 
\item{} Hugin's proper time is increasing from $t_0$ to infinity\footnote{Since Hugin and Munin are mythological creatures we will assume that they are immortal and potentially live forever.}. 
\item{} There is no future singularity. 
\item{} The models are future geodesically complete.
\end{itemize}  
Munin's perspective, on the other hand, depends very much on the value
of the equation of state parameter $\gamma$. Results of calculations using boost formulae and an asymptotic analysis show that \cite{CHL-long}: 
\begin{itemize}
\item{} For $2/3<\gamma\leq 1$, the world-lines asymptotically
  approaches those of Hugin's. 
\item{} For $10/9<\gamma<4/3$, Munin experiences a quintessential
  behaviour in the sense that the deceleration parameter obeys $-1<q_{\text{Munin}}<0$.
\item{} The world-lines are not future complete for $4/3<\gamma<2$; the Hubble scalar, shear and acceleration diverge while the proper time is finite! 
\item{} For $6/5<\gamma<2$, the components of the Weyl tensor diverge.
\end{itemize}
It is clear that Hugin and Munin's record of events can be drastically different. The information they collect about the state of our universe seems to differ radically. For example, Hugin sees an ever-expanding ever-lasting universe while Munin, on the other hand, sees a universe that blows up in finite time and ends in a violent 'big rip' singularity. 
It is important to emphasize that Munin experiences dynamical behaviour similar to that of quintessence \cite{quint} or  phantom cosmology \cite{phantom} but without any exotic matter (i.e., unknown matter with postulated equations of state). 

One might be tempted to dismiss these models since observations show that the
Universe is currently evolving into a dark energy-dominated
epoch. However, regardless of whether the matter dominates at present, there will always be relics of other matter
components from earlier epochs. For example, the Universe will always
possess a non-zero radiation component which might become more and more
diluted but is nonetheless non-zero. The same applies to fluids
with a stiffer equation of state, such as certain scalar fields if
they exist. Consider, therefore, a model having a cosmological constant and a
tilted fluid, which is a good model for our Universe given the  current observation of a
 cosmic acceleration. For these models a similar thing happens: Hugin's Universe approaches de Sitter, while Munin's Universe, for $4/3<\gamma<2$, expands so quickly that it ends in a 'big rip' after a finite time. 

This choice therefore has consequences for everything which we observe. Only true invariants will be completely independent of the choice of congruence. Examples of observer dependent quantities include the Hubble parameter, the shear, etc. This is of particular importance in interpreting cosmological data such as, for example, cosmic acceleration from supernovae data \cite{obs}. This may also be relevant to calculations of the cosmic microwave background radiation, which may be used to constrain such effects. Indeed, the effects may potentially be important in resolving the quadrupole anomaly \cite{CMB}.

\bibliographystyle{amsplain}

\end{document}